# Conceptions et usages des plates-formes de formation

▶ **Sébastien GEORGE (ICTT, INSA de Lyon),**
**Alain DERYCKE (TRIGONE, Université de Lille 1)**

Les plates-formes de formation sont au cœur du développement de la formation en ligne. Composantes à part entière des dispositifs de formation, on ne peut les réduire aux seuls aspects technologiques. En effet, les modèles sous-jacents conditionnent la formation dans son ensemble, de la préparation des enseignements jusqu'au déroulement des sessions d'apprentissage. Les recherches liées à ces plates-formes sont nombreuses et leurs enjeux sont importants. C'est ce qui nous a incité à lancer un appel à numéro spécial sur le thème « Conceptions et usages des plates-formes de formation ».

Une plate-forme de formation est un « système informatique destiné à automatiser les diverses fonctions relatives à l'organisation des cours, à la gestion de leur contenu, au suivi des progrès des participants et à la supervision des personnes responsables des différentes formations » (Office de la langue française, 2005). L'équivalent anglais est *Learning Management System* (LMS). Notons qu'actuellement les activités pédagogiques sont davantage mises en avant, ce qui nous conduit à compléter la définition ci-dessous : une plate-forme peut être vue comme un système qui permet de gérer et de donner accès à un ensemble d'activités et de ressources pédagogiques.

À l'origine, ce numéro spécial avait pour objectif, d'une part, de discuter des utilisations et expériences des plates-formes éducatives et, d'autre part, de proposer des technologies ou techniques de conception et de développement en lien avec les plates-formes. Nous avons constaté, à la réception des articles, que peu de soumissions concernaient les utilisations des plates-formes éducatives et que la plupart avaient une approche assez technologique. De ce fait, la coloration donnée à ce numéro spécial est plus informatique que ce qui avait été escompté à l'origine.

Au total, nous avons reçu 24 contributions. Après deux phases de relecture, 9 articles et 2 rubriques ont finalement été retenus. Nous avons choisi de découper ce numéro en 4 thèmes qui reflètent d'une certaine manière l'organisation du domaine telle que nous le percevons :

(1) ingénierie des plates-formes : approche dirigée par les modèles (3 articles) ;

(2) indexation des objets pédagogiques (3 articles et 1 rubrique) ;

(3) analyse de traces d'activités dans les plates-formes (1 article) ;

(4) supports pour l'échange et l'interopérabilité au sein d'une communauté de recherche en EIAH (2 articles et 1 rubrique).

Le premier thème concerne l'architecture et l'ingénierie des plates-formes. Les problèmes posés touchent directement aux situations/scénarios pédagogiques et à leur mise en place : comment faire en sorte que les plates-formes de formation soutiennent au mieux la modélisation des activités pédagogiques ? Dans les contributions retenues pour ce numéro spécial, nous constatons la prédominance d'approches venant du domaine du génie logiciel. Plus précisément, les approches orientées modèles semblent retenir l'attention des chercheurs qui s'attaquent aux problématiques liées aux scénarios pédagogiques : modèles pour leur conception, modèle pour leur représentation, pour leur mise en œuvre, etc.

Le deuxième thème de recherche, complémentaire du précédent, est lié aux ressources/objets pédagogiques et s'intéresse aux objets informatiques qui décrivent les contenus et connaissances d'un domaine à enseigner. Les questions qui se posent au sujet des objets pédagogiques sont nombreuses : conception, structuration, description, indexation, partage, recherche, réutilisation, assemblage, intégration, etc. Les contributions de ce numéro dans ce thème abordent essentiellement les modèles de description et la recherche des objets pédagogiques.

Le troisième thème concerne l'utilisation des plates-formes. Les problématiques, souvent pluridisciplinaires, touchent à l'analyse des processus de mise en œuvre de plates-formes, à l'étude des impacts des plates-formes sur les contenus et sur l'organisation des formations et à l'observation des utilisations réelles que font les différents acteurs des plates-formes. Les questions sous-jacentes sont celles des méthodes et des outils pour recueillir et analyser les informations sur les activités qui se déroulent. Nous avons reçu peu de contributions directement dans ce thème, même s'il est vrai que la plupart des recherches du domaine abor-dent des questions d'utilisation afin de valider leurs modèles ou comme méthode de conception. Au final, un article se situe dans le thème de l'utilisation et aborde une question majeure, celle de l'analyse des traces d'activités d'étudiants.

Enfin, le quatrième thème est d'une autre nature. Il s'agit du support à la communauté de recherche en EIAH. Les EIAH produits ont, pendant longtemps, été conçus et développés de manière isolée, pour des besoins spécifiques. Le domaine de recherche des EIAH souffre parfois d'une « difficulté à capitaliser les ''briques'' qui sont ici et là construites à la faveur d'une thèse ou d'un projet » (RTP39, 2002). Ainsi, l'efficacité de la recherche en EIAH est faible en termes de mutualisation et de capitalisation des objets techniques produits. Cette considération ne doit pas rester à la marge mais doit devenir une préoccupation centrale car elle conditionnera la capacité de la communauté à développer et à conforter son activité scientifique. La question centrale est de voir comment il est possible de favoriser la mutualisation des outils, composants et prototypes logiciels développés par la communauté de recherche.

### 1. Ingénierie des plates-formes : approche dirigée par les modèles

Les premiers travaux sur les plates-formes éducatives se sont concentrés sur les ressources à mettre à disposition aux apprenants. Depuis, des recherches ont placé en position centrale les activités pédagogiques et non les ressources (Koper, 2000 ; Koper et Tattersall, 2005), sans pour autant négliger les ressources pédagogiques nécessaires à la réalisation des activités (voir thème 2).

Rob Koper, de l'Université Ouverte des Pays-Bas, s'est démarqué en s'intéressant à la modélisation pédagogique pour mieux répondre à l'intégration effective des technologies dans la formation. Il a conçu un langage de modélisation pédagogique nommé EML (*Educational Modelling Language*) qui est à l'origine du standard adopté par IMS en 2003 : *IMS-Learning Design* (IMS-LD, 2003). Dans IMS-LD, le scénario pédagogique est décrit comme une mise en scène (métaphore théâtrale). Une pièce est composée d'actes, eux mêmes composés de partitions qui associent un rôle à une activité effectuée dans un environnement. Un acteur peut alors être un apprenant ou un enseignant qui encadre l'activité.

Selon Guéraud *et al.* (2004), « un scénario pédagogique est défini par :
– la situation initiale et l'objectif à atteindre ;

– les situations correspondant aux étapes de résolution pertinentes ;

– les situations particulières à observer (contraintes à respecter, erreurs classiques, dangers potentiels,…) ;

– la réactivité permettant d'assister l'apprenant en fonction de sa progression ; elle détermine les réactions du système (retours d'information, aide, retour au début d'étape,…) associées aux différents contrôles (étape réussie ou non, situation particulière atteinte, objectif atteint ou non). »

Dans ce numéro spécial, trois articles s'intéressent au support de la modélisation pédagogique dans les plates-formes de formation. Ces trois contributions adoptent toutes une approche génie logiciel et même, plus précisément, une Ingénierie Dirigée par les Modèles (IDM, ou MDE pour *Model Driven Engineering*). « L'IDM se distingue des méthodes de modélisations traditionnelles […] par la préoccupation constante de rendre les (meta) modèles productifs plutôt que contemplatifs » (Bézivin *et al.*, 2004). Le lecteur pourra se référer à des travaux entièrement dédiés à l'IDM (ASMDA, 2006 ; Schmidt, 2006) ou à l'IDM pour les EIAH (Nodenot, 2005). Les contributions de ce numéro abordent les questions suivantes : le support à l'ingénieur pédagogique ou, de manière générale, à l'équipe de conception, les langages semi-formels de modélisation de scénarios pédagogiques et le support à l'exécution de scénarios. Voici un bref résumé de leurs apports.

**Corbière et Choquet** partent de résultats provenant de la communauté génie logiciel pour définir un cadre visant à supporter une communauté de concepteurs d'un système de formation. Le modèle ODP-RM (*Open Distributed Processing – Reference Model*), cadre de développement proposé par l'ISO, est alors interprété dans le contexte de l'ingénierie des EIAH dans l'objectif de formaliser la communication entre concepteurs de tels systèmes. Les normes des technologies éducatives (IMS-LD, LOM, SCORM entre autres) sont abordées selon un point de vue rétroconception et réingénierie, mettant ainsi en avant l'aspect évolutif de tout système de formation. Les auteurs illustrent le modèle proposé au travers de deux utilisations : la rétroconception d'un EIAH, c'est-à-dire le fait de le spécifier avec plus de détails ; la réingénierie d'un EIAH, pour guider des concepteurs dans les modifications à effectuer (démarche qualité). Dans les deux cas, le cadre ODP-RM aide à mieux expliciter les systèmes dans le but fédérer des expériences de formation.

Suivant également une approche issue du génie logiciel, **Laforcade, Nodenot et Sallaberry** proposent une représentation graphique pour la modélisation des situations pédagogiques. Ils présentent un langage nommé CPM (*Coopérative Problem-based learning Metamodel*), spécialisation d'UML, encore appelé « profil UML », orienté vers la conception de situations-problèmes coopératives. La cible première est l'ingénieur pédagogique lors de l'étape d'expression initiale des besoins. Ce langage CPM a été implanté dans un atelier de génie logiciel (*objecteering/UML*). Une expérimentation du prototype conçu a permis de valider l'approche qui serait exploitable pour d'autres types de situations pédagogiques. La vocation principale du langage graphique CPM est de faciliter la communication dans les phases initiales de conception d'une situation pédagogique, mais il n'y a pas encore de méthode associée qui guiderait la conception. Une perspective intéressante est alors de voir comment relier ce type de langage à d'autres langages plus universels comme IMS-LD se situant dans des phases plus aval de la conception. Pour cela, les auteurs proposent une première ébauche consistant à générer automatiquement un fichier XML conforme à la spécification IMS-LD à partir d'un diagramme d'activité décrit avec CPM. C'est un premier pas vers une méthode d'ingénierie pédagogique qui serait dirigée par des modèles.

Les travaux de **Vantroys et Peter** se situent également dans le domaine de l'ingénierie des EIAH avec une approche fortement orientée modèle. Leur article commence par une description précise du concept de scénario pédagogique, en termes de conception et d'utilisation. Pour compléter cette description, les rôles que l'on peut rencontrer dans une plate-forme de formation font l'objet d'une classification. Selon les auteurs, les plates-formes actuelles se focalisent trop sur l'accès aux ressources et pas assez sur la scénarisation des activités. Ils proposent alors un système pour supporter le plus complètement possible les scénarios pédagogiques dans les plates-formes, notamment leur exécution. Une implémentation du système a été réalisée sous forme de service web que toute plate-forme peut utiliser. Le système repose sur un moteur de *workflow*, nommé COW (*Cooperative Open Workflow*), qui a été réalisé en respectant le standard WMF (*Workflow Managment Facility*) de l'OMG et langage standard de spécification externe des flots de travail XPDL. Pour valider l'approche, la connexion de COW avec la plate-forme « Campus Virtuel » a été réalisée avec succès. Pour le moment COW fonctionne avec un langage de modélisation de scénarios pédagogiques spécifique (COW *Language*), qui est ensuite traduit en XPDL, même si un futur support de IMS-LD est prévu.

Le schéma suivant (figure 1) synthétise les contributions précédentes, avec une cartographie simplifiée permettant de les positionner les unes par rapport aux autres. Ce schéma a été réalisé en utilisant la technique de modélisation par objets typés MOT (Paquette, 2002).

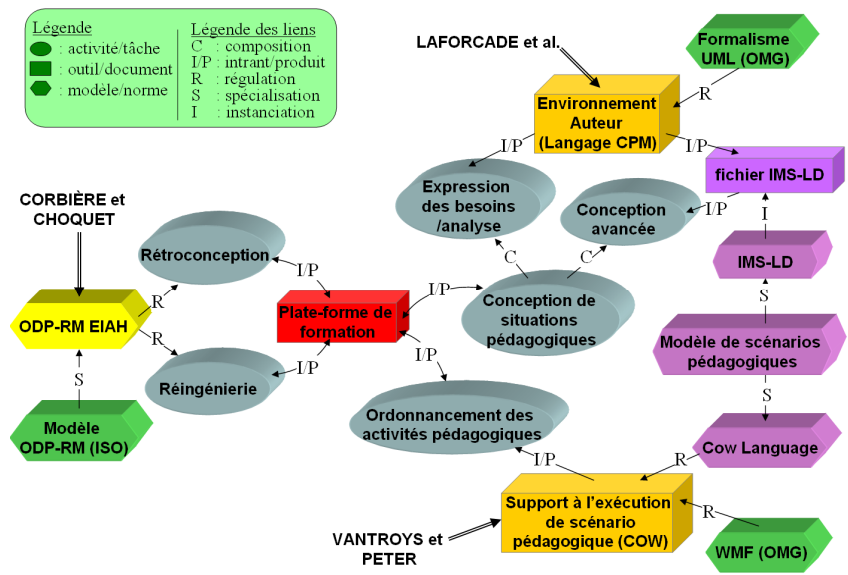

**Figure 1 • Synthèse des contributions sur le thème
« Ingénierie des plates-formes : approches génie logiciel »**

## 2. Indexation des objets pédagogiques

Selon l'IEEE, un objet pédagogique ou objet d'apprentissage (*Learning Object*) est une entité sur support informatique ou non, qui peut être utilisée, réutilisée ou référencée dans une activité de formation assistée par ordinateur (IEEE, 2005). Se posent alors des questions autour des modalités de structuration des objets pédagogiques et de description de leur contenu (métadonnées). Les enjeux concernent la réutilisation des objets pédagogiques et la possibilité de fournir des apprentissages individualisés.

Les travaux sur la description de contenus ont donné lieu à la définition de schémas de métadonnées pédagogiques. Un standard international a alors émergé, LOM (*Learning Object Metadata*), proposant un modèle de description des métadonnées associées à des objets pédagogiques. LOM est défini comme étant « *the attributes required to fully/adequately describe a Learning Object* » (IEEE, 2005). Cependant, LOM ne convient pas forcément directement à tous les contextes d'utilisation. Il est alors possible de définir des profils d'application du LOM, un profil d'application étant une

« personnalisation d'une norme pour répondre à des communautés particulières de réalisateurs ayant des exigences communes en matière d'applications » (Lynch, 1997). De manière plus précise, un profil d'application est « assemblage d'éléments de métadonnées choisis parmi un ou plusieurs schémas de métadonnées et combinés pour former un schéma composé. Les profils d'application permettent de mettre en œuvre les principes de modularité et d'extension. L'objectif d'un profil d'application est d'adapter ou de combiner des schémas existants afin d'obtenir un nouveau schéma conçu pour une application particulière tout en gardant l'interopérabilité avec le ou les schémas de base. Cette adaptation peut inclure la définition d'éléments de métadonnées locaux, qui sont importants pour une communauté mais qui ne le sont pas dans un contexte plus large », traduction de l'AFNOR (2005) de Duval *et al.* (2002). Par exemple, dans le cadre d'une convention liant le ministère de l'éducation nationale à l'AFNOR, un profil d'application français du LOM (LOM.fr) a vu le jour (publié à titre expérimental en août 2005). Le profil ManUel, conçu pour le campus numérique C@mpuSciences® (De La Passardière et Jarraud, 2004), constitue un autre exemple français.

Lorsque l'on aborde la question de la réutilisation des objets pédagogiques d'un point de vue technique, on parle alors d'interopérabilité, qui peut se définir comme étant « le caractère de ce qui permet l'utilisation des ressources d'enseignement et d'apprentissage développées par une organisation dans un environnement technologique donné par d'autres organisations dans d'autres environnements technologiques » (Normetic, 2005). Ainsi, un élément référencé dans une représentation doit pouvoir aussi être référencé dans une autre représentation (correspondance possible). Nous retrouvons à ce niveau les viviers de connaissances ou LOR (*Learning Object Repository*), lieux de stockage des objets pédagogiques.

**Bourda et Delestre** cherchent à améliorer l'interopérabilité entre différents profils d'application du LOM, en définissant un métamodèle afin de favoriser la description standardisée des données utilisées dans les différents profils. Le métamodèle proposé est fondé sur des concepts et non sur des termes. Cette solution a pour objectif d'éviter les problèmes d'interopérabilité actuelle : mettre sous le même nom de concepts distincts ou bien utiliser des noms différents pour un même concept. La séparation de l'aspect conceptuel et de l'aspect représentation, préconisée par la norme ISO11179, est la base du schéma de métadonnées conçu. Les auteurs décrivent alors une architecture en 4 niveaux, reposant sur des transformations successives de fichiers XML, partant de la modélisation conceptuelle du domaine à représenter et allant jusqu'à l'instanciation pour un contexte donné. Cette architecture a été développée en partie et devrait aboutir à terme sur un outil d'aide à la conception de schémas de métadonnées compatibles avec l'ISO11179.

Dans leur article, **Broisin et Vidal** s'intéressent à la virtualisation des objets pédagogiques. Cette virtualisation consiste à proposer une interface fournissant une vue unifiée sur des objets se trouvant dans des viviers de connaissance distincts et ce de manière transparente pour l'utilisateur. Différents viviers de connaissance peuvent ainsi être fédérés, augmentant le nombre de ressources disponibles à partir d'un outil unique. L'objectif est de mettre à disposition un ensemble de ressources aux utilisateurs tout en leur fournissant des services associés (recherche, accès, consultation, …). En particulier, un service offre la possibilité de générer automatiquement un certain nombre des métadonnées, comme celles de LOM, nécessaires à l'indexation des objets pédagogiques. Pour finir, deux expérimentations sont détaillées afin de valider l'architecture de virtualisation. Une interface de recherche d'objets pédagogiques, prenant la forme d'un moteur de recherche, offre à l'utilisateur un accès à des ressources provenant de trois viviers. La couche d'abstraction proposée a été connectée avec succès à deux plates-formes d'apprentissage différentes.

**Bouzeghoub**, **Defude**, **Duitama et Lecocq** tentent, quant à eux, d'étendre les standards actuels en leur adjoignant un modèle de description sémantique des ressources pédagogiques, fondé sur une ontologie de domaine. Ainsi, en plus des caractéristiques définies par LOM, une ressource est représentée par les pré-requis nécessaires, son contenu et une fonction d'acquisition permettant de modifier le modèle de l'apprenant. Cette description permet d'offrir des outils de recherche et de composition plus élaborés. Une implémentation de ce modèle a été réalisée sous la forme d'une base de données RDF qui dispose d'un langage de requête. Des tests d'utilisation sont actuellement en cours.

La figure 2 donne une cartographie très générale pour situer les trois contributions sur le thème de l'indexation des objets pédagogiques.

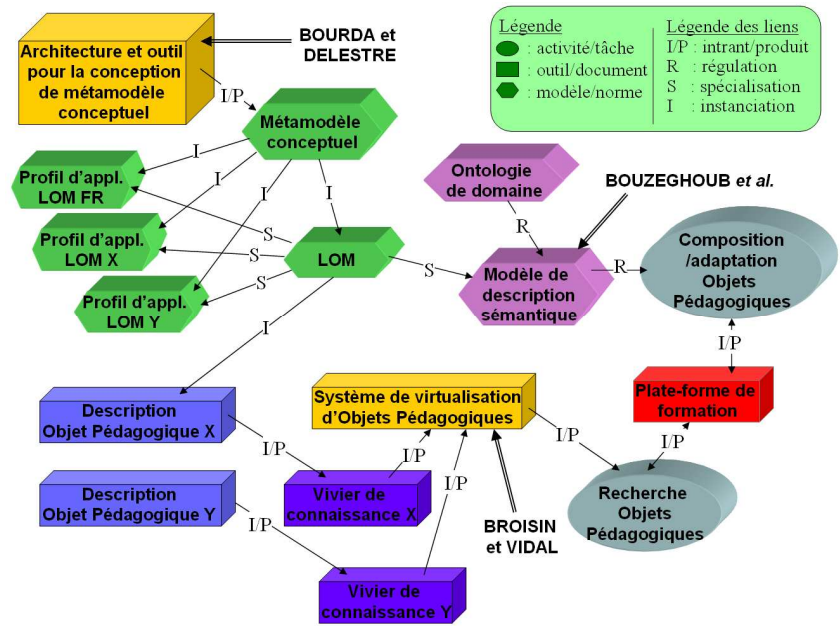

**Figure 2 • Synthèse des contributions sur le thème
« Indexation des objets pédagogiques »**

Dans leur rubrique, **Gebers**, **Campana**, **Ettori**, **Papi et Majada** décrivent une architecture pour séparer les phases de production de contenu (avec un LCMS, *Learning Content Management System*) et celles de diffusion (avec un LMS), ce qui peut s'avérer particulièrement utile dans un contexte d'industrialisation de la formation. Cette approche présente un intérêt certain au plan organisationnel mais peut conduire à des problèmes d'interopérabilité, nécessitant le recours à l'utilisation de standards. Les auteurs décrivent alors la mise en œuvre du passage au standard SCORM dans leur dispositif de formation, et en ressortent les inconvénients (perte de fonctionnalités,…) et les avantages (meilleur suivi des apprenants). En conclusion, l'architecture proposée, dissociant LCMS et LMS, montre sa capacité à bénéficier rapidement de l'amélioration des standards.

### 3. Analyse de traces d'activités dans les plates-formes

Toutes les plates-formes de formation possèdent des fonctionnalités de recueil et de stockage d'information sur les activités des différents acteurs. Ces « traces d'activités » peuvent être plus ou moins fines, allant du sim-ple fichier de *logs* contenant des informations sur les connexions des utilisateurs jusqu'aux traces contenant toutes les interactions des utilisateurs avec le système. L'exploitation de ces traces recueillies demeure une problématique centrale dans les recherches autour des plates-formes.

Les travaux de **Vanderbrouk et Cazes** concernent l'exploitation de fichiers traces ou *logs* contenant des informations sur les actions des utilisateurs dans une plate-forme d'enseignement dans le domaine des mathématiques. Un modèle didactique permet de passer d'une information quantitative à une analyse qualitative, en vue d'obtenir des renseignements sur le fonctionnement didactique des étudiants lors de la réalisation d'exercices. À l'aide de deux indices calculés d'après les activités effectivement réalisées par les groupes d'étudiants, l'indice de difficulté d'un exercice et le rendement (rapport entre la note et le temps passé sur un exercice), les auteurs déterminent alors 3 classes d'exercice (simples, intermédiaires et difficiles). Cette catégorisation permet par exemple de fournir des informations pour re-concevoir les scénarios ou prévoir l'encadrement afin de mieux atteindre les objectifs pédagogiques visés.

### 4. Supports pour l'échange et l'interopérabilité au sein d'une communauté de recherche en EIAH

Une préoccupation actuelle dans le domaine des EIAH concerne la réutilisation des produits de recherche développés. En particulier, un regroupement international au sein du Réseau d'Excellence Kaléidoscope (2005), le sous-projet SVL (*Shared Virtual Laboratory*), vise à développer une plateforme technologique de services pour la communauté de recherche en EIAH (Kaleidoscope, 2005 ; SVL, 2005). Nous présentons dans cette partie deux articles et une rubrique sur ce thème.

L'article de **Rosselle, Bessagnet et Carron** part du constat que peu de logiciels EIAH arrivent réellement à sortir du contexte expérimental des laboratoires pour aller vers une utilisation effective dans les classes. Une des raisons soulevées ici est qu'un EIAH ne réalise souvent qu'une partie des fonctionnalités nécessaires pour que l'utilisation soit efficace et réellement intégrée dans les pratiques courantes. L'objectif est alors de proposer des solutions techniques favorisant l'intégration de fonctionnalités provenant d'EIAH différents, correspondant à la problématique globale de l'intégration de composants hétérogènes. L'article se centre sur les spécificités particulières des EIAH en insistant sur les questions à se poser selon certaines dimensions : indexation, échange de données, IHM, etc. Pour illustrer l'intégration d'EIAH, des études de cas sont présentées. En particulier, deux intégrations d'EIAH sont analysées en détail : l'une porte sur l'utilisation conjointe de fonctionnalités provenant de deux EIAH et l'autre sur la composition de deux applications EIAH. En conclusion, les auteurs affirment qu'il est encore trop tôt pour définir « un modèle d'intégration générique » et propose plutôt des recommandations pour expérimenter de nouvelles intégrations.

Les travaux de **Rebaï, Maisonneuve et Labat** se situent dans la même lignée. Le constat de départ est identique : il est difficile de construire un environnement EAIH complet. Les nombreuses fonctionnalités nécessaires pour qu'un EIAH soit utilisable impliquent forcément des temps de développement importants. L'idéal serait de pouvoir réutiliser des fonctionnalités existantes, ce qui nécessite par conséquent une description et indexation des composants logiciels EIAH existants afin de faciliter leur recherche. Un premier résultat des auteurs concerne la définition de quatre classes de composants logiciels EIAH : pédagogiques, de services, techniques et enfin de fabrication (outils auteurs). Puis les auteurs montrent les faiblesses des normes et standards actuels pour décrire des composants, notamment pour la description des aspects techniques. Un ensemble de métadonnées visant à combler cette lacune est ainsi proposé (LSCM : *Learning Software Component Metadata*). Les métadonnées sont divisées en deux jeux : un jeu commun à tous les composants logiciels et un jeu spécifique en fonction des 4 classes de composants définies ci-dessus. Enfin, les auteurs décrivent la conception d'un prototype de plate-forme de mutualisation destinée aux chercheurs en EIAH : *Educational Component Repository* (ECR). Cette plate-forme est qualifiée de générique car elle prend la forme d'un entrepôt de composants logiciels acceptant plusieurs schémas de description (métadonnées). Des expérimentations à venir de la plate-forme devraient pouvoir confirmer son utilité pour la communauté EIAH.

Dans une rubrique, le collectif **AS Plateforme pour la recherche en EIAH** décrit sa contribution à l'ingénierie des EIAH. Les résultats présentés sont de trois natures. D'abord une définition de scénarios d'usage visant à déterminer les besoins potentiels de la communauté de recherche en EIAH. Puis l'étude technique d'un portail pour la communauté, de type *open source*. Enfin, l'identification d'une banque de scénarios pédagogiques de référence. En vue d'améliorer la diffusion et la réutilisation des résultats de la recherche au sein de la communauté de recherche en EIAH, le projet devrait pouvoir déboucher sur la création d'un portail dédié à la communauté de recherche francophone en EIAH, portail dérivé du projet SVL du réseau d'excellence Kaléidoscope.

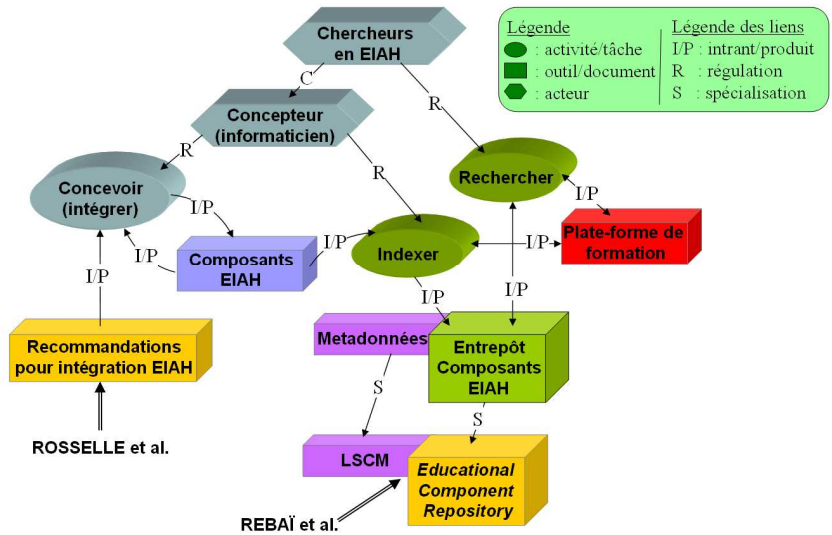

**Figure 3 • Synthèse des contributions sur le thème
« Architecture pour l'intégration d'EIAH/approche composant »**

## 5. Conclusion

Les travaux présentés dans ce numéro spécial de STICEF consacré aux plates-formes pour l'apprentissage fait apparaître plusieurs éléments intéressants :

– il y a une évolution des travaux de recherche sur les EIAH vers une approche plus macroscopique incluant le dispositif de formation dans sa globalité et mettant l'activité de l'apprenant plus en valeur au travers de la modélisation des scénarios pédagogiques. Cette évolution est sans doute induite par le rapide développement de l'e-formation utilisant le support de l'Internet ;

– la communauté de recherche s'est tournée résolument vers le domaine du Génie Logiciel pour lui donner une approche plus solide où les préoccupations de réutilisabilité et d'interopérabilité des environnements EIAH sont devenues plus présentes ;

– il y a un souci plus grand également de fournir des méthodes, et des instruments afin de favoriser la collaboration au sein de la communauté de recherche en EIAH.

C'est l'une des retombées de l'action collective menée ces dernières années, notamment au sein du Réseau Thématique Pluridisciplinaire « Apprentissage, Formation, Éducation » et de son action spécifique, présentée ici dans la section rubrique. Cette action a également fait l'objet d'une dissémination plus large au travers d'une école thématique du CNRS qui s'est déroulée en juillet 2005 (ÉcoleEIAH, 2005).

Il faut espérer que ce numéro spécial va renforcer cette volonté de collaboration et de mutualisation et faire prendre conscience, notamment aux décideurs et tutelles diverses, que cet effort doit être soutenu au niveau national et au niveau international, notamment de la francophonie.

## Références à des sites Internet